\documentclass[prl,twocolumn,showpacs,superscriptaddress,preprintnumbers,floatfix]{revtex4-1}

\usepackage{etex}
\usepackage{ifpdf}
\usepackage{hyperref}
\usepackage{dcolumn}
\usepackage{url}
\usepackage{amsmath}
\usepackage{amscd}
\usepackage{amsfonts}
\usepackage{amssymb}
\usepackage{amsthm}
\usepackage{xfrac}
\usepackage{braket}
\usepackage{bm}   
\usepackage{verbatim}
\usepackage{stmaryrd}
\usepackage{xcolor}
\usepackage{setspace}
\usepackage{tikz}
\usetikzlibrary{arrows}
\usetikzlibrary{automata}
\usetikzlibrary{positioning}
\usetikzlibrary{plotmarks}
\usepackage{pgfplots}
\pgfplotsset{compat=newest}
\usepackage{floatrow}

\tikzstyle{vaucanson}=[
  node distance=2.5cm,
  bend angle=15,
  auto,
  every loop/.style={},
  every edge/.style={->,draw=black,line width=1.2,>=latex,shorten <=1pt,shorten >=1pt},
  every state/.style={draw=black,line width=2,font=\large},
  loop right/.style={right,out=22,in=-22,loop},
  loop above/.style={above,out=112,in=68,loop},
  loop left/.style={left,out=202,in=158,loop},
  loop below/.style={below,out=292,in=248,loop},
  loop above right/.style={right,out=67,in=23,loop},
  loop above left/.style={left,out=157,in=113,loop},
  loop below left/.style={left,out=247,in=203,loop},
  loop below right/.style={right,out=337,in=293,loop},
]

\newcommand{\Prob}{\mathrm{Pr}}

\theoremstyle{plain}    
\theoremstyle{plain}    
\theoremstyle{plain}    
\theoremstyle{plain}    
\theoremstyle{plain}    
\theoremstyle{plain}    
\theoremstyle{plain}    
\theoremstyle{plain}    
\theoremstyle{plain}    
\theoremstyle{plain}    
\theoremstyle{plain}    
\theoremstyle{plain}    
\theoremstyle{plain}    

\colorlet {R_color}    {blue}
\colorlet {k_color}    {black!30!green}

\def\clap#1{\hbox to 0pt{\hss#1\hss}}

\begin{document}

\title{Weak universality in sensory tradeoffs}

\author{Sarah Marzen}
\email{smarzen@berkeley.edu}
\affiliation{Department of Physics,\\
Redwood Center for Theoretical Neuroscience\\
University of California at Berkeley, Berkeley, CA 94720, USA}

\author{Simon DeDeo}
\affiliation{Center for Complex Networks and Systems Research, Department of Informatics, Indiana University, 919 E 10th St, Bloomington, IN 47408, USA}

\date{\today}
\bibliographystyle{unsrt}

\begin{abstract}
{
For many organisms, the number of sensory neurons is largely determined during development, before strong environmental cues are present. This is despite the fact that environments can fluctuate drastically both from generation to generation and within an organism's lifetime. How can organisms get by by hard-coding the number of sensory neurons?  We approach this question using rate-distortion theory.  A combination of simulation and theory suggests that when environments are large, the rate-distortion function---a proxy for material costs, timing delays, and energy requirements---depends only on coarse-grained environmental statistics that are expected to change on evolutionary, rather than ontogenetic, timescales.
}

\vspace{0.2in}

\end{abstract}

%

\maketitle


\setstretch{1.1}

\newcommand{\Abet}{\ProcessAlphabet}
\newcommand{\MS}{\MeasSymbol}
\newcommand{\ms}{\meassymbol}
\newcommand{\SSet}{\CausalStateSet}
\newcommand{\St}{\CausalState}
\newcommand{\st}{\causalstate}
\newcommand{\FSt}{\FutureCausalState}
\newcommand{\fst}{\futurecausalstate}
\newcommand{\FCmu}{\FutureCmu}
\newcommand{\PCmu}{\PastCmu}
\newcommand{\PSt}{\PastCausalState}
\newcommand{\pst}{\pastcausalstate}
\newcommand{\MxSt}{\AlternateState}
\newcommand{\MxSSet}{\AlternateStateSet}
\newcommand{\mxst}{\mu}
\newcommand{\mxstt}[1]{\mu_{#1}}
\newcommand{\StartMS}{\bra{\delta_\pi}}

\newcommand{\CodeRate}  { \I [\Past;\AlternateState] }
\newcommand{\Shielding} { \I [\Past;\Future | \AlternateState] }
\newcommand{\StateFutI} { \I [\AlternateState | \Future ] }


The amount of sensory information potentially available to an organism is, for all practical purposes, infinite. This, taken together with the finite size of the brain, implies that
we constantly operate in the lossy regime, transmitting only some of the information present in the environment.  Optimal sensing, in other words, is optimal compression, and this means that core theorems of information theory constrain biology of perception.
 In the context of evolved sensors, the introduction of the need to compress and coarse-grain environmental signals extends the efficient coding hypothesis~\cite{barlow}, which has guided experimental and theoretical neuroscience for the past five and a half decades \cite[and references therein]{simoncelli2001natural}, to the lossy regime \cite{salisbury2015optimal}.
 
Considering the role of lossy compression in evolved sensory systems leads to interesting interpretations of existing experimental results concerning neurogenesis, or the dynamic creation of new neurons over an organism's lifetime~\cite{gross2000neurogenesis,paridaen2014neurogenesis,green2014inflammation,kirste2015silence}. While neurogenesis is widespread, neurogenesis in sensory regions is less commonly observed.
Indeed, the number of neurons in sensory regions appears to be essentially determined prior to receipt of any environmental cues~\cite{williams1988control}, though some famous counterexamples exist~\cite{kaslin2008proliferation}.  In other words, for many species, the number of neurons in a brain's sensory region is strongly determined by fixed, genetic effects~\cite{Lallemend2012373}, even if this process continues late into development~\cite{kaslin2008proliferation}.


Here, we provide an information-theoretic explanation for these facts by viewing early sensory regions as lossy perceptual feature extractors for which the number of sensory neurons limits the accuracy of the organism's internal representation of the environment.  To show this, we use a model of the environment general enough to apply to a range of biological situations, but rich enough to capture the basic problem of perception and encoding, in which both the probability of observing a particular environmental symbol and the cost of misperceiving those symbols are randomly drawn~\cite{marzen2015evolution}.

In this minimal model, the tradeoff between neuron number and representational accuracy is essentially invariant to changes in the probability distribution over sensory inputs and the particular costs of misperceiving one sensory input for another; this is true even though the optimal internal coding of environmental inputs varies wildly from one environment to the next.  These results lead to a new functional interpretation of phenotypic variability and neurogenesis in sensory brain regions: first, phenotypic variability in sensory neuron number may be tied to phenotypic variability in the average heat dissipation rate of a sensory neuron; and second, neurogenesis may only be necessary when the organism-environment interactions change drastically, \emph{e.g.}, due to changes in action policy.

\section{Relating sensory costs to the rate-distortion function}
\label{sec:Costs}

Confusing one environmental state for another can be costly due to a subsequent suboptimal choice of action.  For example, mistaking a lion for a domesticated cat might lead to death, while mistaking a domesticated cat for a lion might lead to unnecessary energy spent running. However, correctly identifying a greater number of objects requires more mental effort, whether that be measured by a larger number of neurons devoted to object recognition or a correspondingly larger number of ATP molecules consumed in their function.
Previous work suggests that resource constraints such as these are critical in shaping the neural code \cite{laughlin2001energy,varshney2006optimal,balasubramanian2001metabolically,laughlin1998metabolic,levy1996energy,schreiber2002energy,chklovskii2004maps}.

An optimal sensor uses as few resources as possible to achieve a desired accuracy. Rate-distortion theory, a branch of information theory that deals with \emph{lossy} communication, places asymptotically achievable lower bounds on the \emph{rate} of the sensor, which is the bits per input symbol required to communicate the sensor's state to a decoder.  The sensor's \emph{distortion} is given by the expected value of a user-specified distortion measure $d(x,\tilde{x})$, which measures the cost of confusing $x$ and $\tilde{x}$. Distortion can be connected to the reward function $r(x,a)$ and action policy $p(a|x)$ in a simplified reinforcement learning setup \cite{barto1998reinforcement} where $a$ are possible actions, via $d(x,\tilde{x}) = \left(\max_{\tilde{x}} \sum_{a} p(a|\tilde{x}) r(x,a)\right) - \sum_{a} p(a|\tilde{x}) r(x,a)$.  This distortion measure is ``normal'', $d(x,x)=0$, if the action policy uses all available information about the environmental state via sensory representation.

\begin{figure}
\centering
\includegraphics[width=\textwidth]{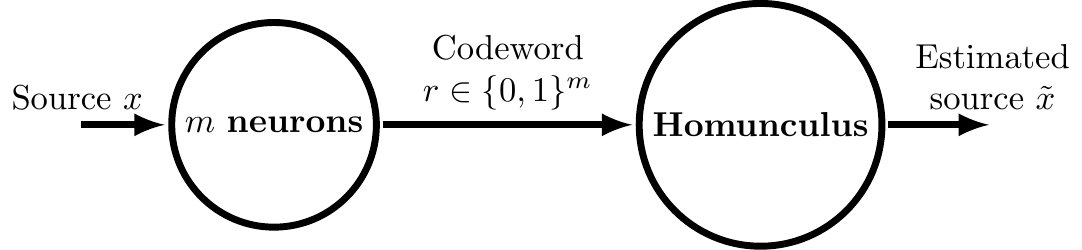}
\includegraphics[width=\textwidth]{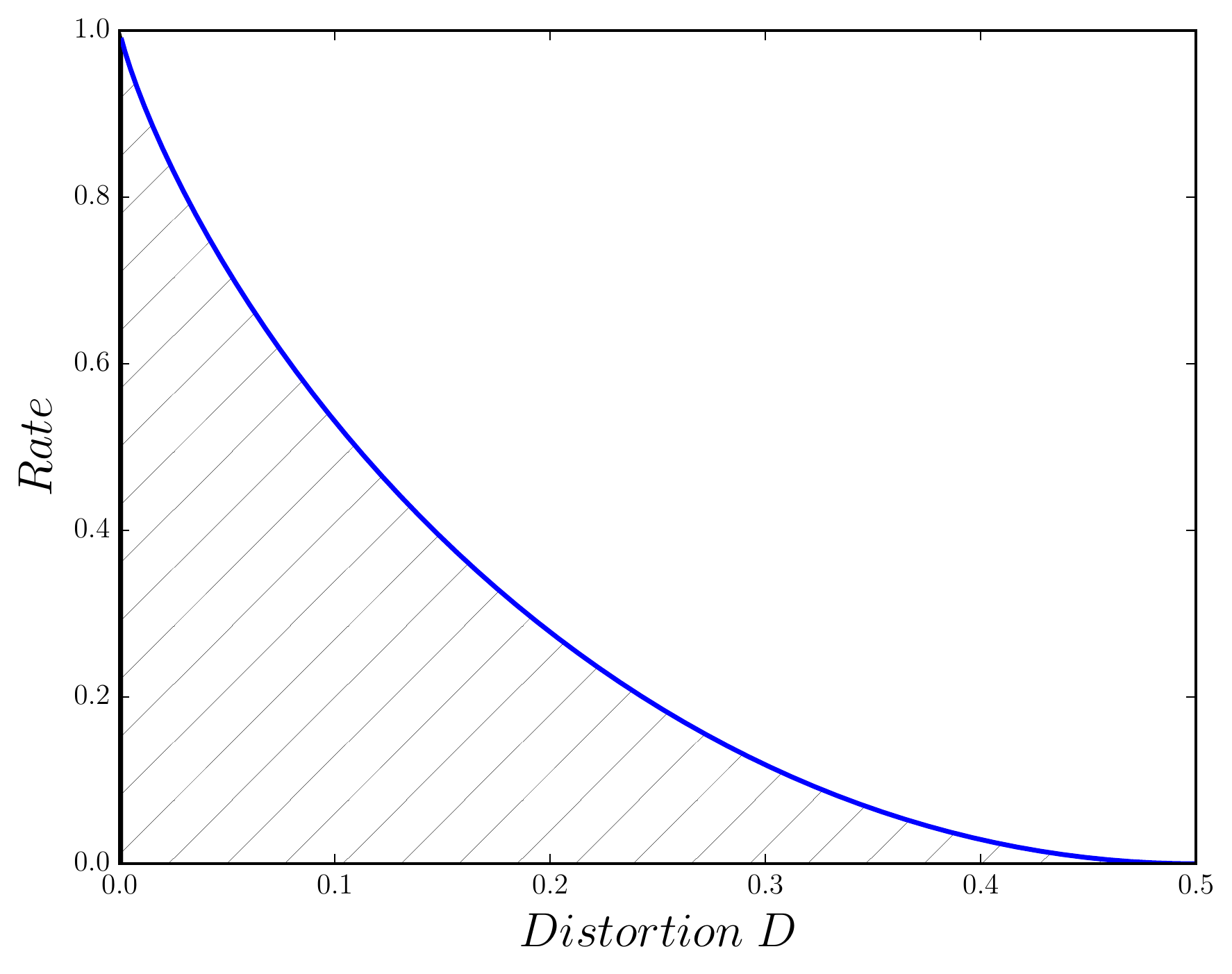}
\caption{\textbf{Rate-distortion theory: estimation with an information bottleneck.}  Top panel: $m$ sensory neurons are asked to communicate an environmental input $x$ to a ``homunculus'' using neuron spike code $p(r|x)$; the homunculus creates an estimate, $\tilde{x}$, of the environmental input on the basis of the spiking pattern $r$.  The quality of estimation is given by the expected distortion $D=\langle d(x,\tilde{x})\rangle_{p(x,\tilde{x})}$, and the absolute rate of the information bottleneck, $R$, is $m$, the number of sensory neurons.  Bottom panel: the rate-distortion function shown in blue delineates the boundary between achievable (white) and unachievable (hatched) combinations of rate and distortion.  The rate-distortion function shown here is that for two equiprobable environmental inputs and Hamming distortion measure \cite{Cove06a}.}
\label{fig:0}
\end{figure}

The rate-distortion function $R(D)$ delineates the boundary between achievable and unachievable combinations of rate and distortion as shown in Fig.~\ref{fig:0}; if the sensor codes $n$ successive input symbols using one codeword, then the lower bound given by the rate-distortion function on the bits required to communicate said codewords per input symbol is achievable in the limit of arbitrarily large $n$, i.e. arbitrarily large delays between sensing and action.

Researchers have used rate-distortion theory to study everything from chemotaxis \cite{iglesias2007} to genetic transcription \cite{tlusty2008} to prediction in the salamander retina \cite{Palm13} to human vision \cite{sims2015}. The appropriate choice of information source and distortion measure depends heavily on the particular biological system that one studies. Here, we use rate-distortion theory to model environments using the framework of Ref.~\cite{marzen2015evolution}. 

Distortions $d(x,\tilde{x})$ and the probability distribution of inputs $p(x)$ are drawn from a probability distribution that represents the range of possible environments an organism might find itself born into. In this paper, for simplicity, all off-diagonal distortions are drawn i.i.d. with probability density function $\rho(d)$, and the probability distribution of inputs is drawn from a Dirichlet distribution with concentration parameter $\alpha$. The $\alpha$ parameter in Dirichlet distribution dictates how uncertain an environment is; as $\alpha$ increases, $p(x)$ is more evenly distributed among the $N$ possible states, and the uncertainty of the environment increases.  Together, $\rho(d)$ and $\alpha$ specify a generative model for environments.


The exact relationship between the rate of a sensor and its ``resource costs''---material costs, power consumption, or timing delays---depends on the physical substrate which lossily communicates environmental information.
For concreteness, consider $m$ sensory neurons that form an information bottleneck between environmental information and downstream brain regions that decide organism actions based on the perceptual information.  This coding distorts the sensory representation with an expected distortion of $D = \langle d(x,\tilde{x}) \rangle_{p(x,\tilde{x})}$.
Though some work suggests that the neural code might be analog (based on spike timing), there is inherent noise in neural circuitry that effectively imposes a minimal discretization time (a few milliseconds \cite{nemenman2008neural}) on which the neural code operates.  We choose the time units to be that minimal discretization time, and think of the sensory neural code as a binary vector of length $m$ in which $1$ ($0$) in the $i^{th}$ position codes for a spike (no spike) from neuron $i$ in that minimal discretization time.

Material and timing costs can trade off with one another, but both run into fundamental limits quantified by the rate-distortion function.  If the number of sensory neurons is greater than the rate-distortion function, $R(D)\leq m$, then one can instantaneously decode each estimated input $\tilde{x}_t$ from the $t^{th}$ binary vector of length $m$, but $R(D)$ places a lower bound on material cost $m$.  If $R(D)\geq m$, then we can acquire additional expressiveness by coding each input $x$ as a string of binary vectors of length $m$, resulting in timing delays.  The expected length of the neuronal output string is no less than $R(D)/\log_2 2^m = R(D)/m$, which, when multiplied by the number of input symbols sensed thus far, is the timing delay between encoding and decoding.  

Finally, a more generally applicable nonequilibrium thermodynamics viewpoint ties the rate-distortion function to power consumption.  Memoryless channels implicitly have a measure-reset cycle: first, the channel senses the environment, and the channel communicates its measurement to some ``homunculus''; and afterwards, the channel resets its internal state. The energy per reset required to maintain such a channel is lower-bounded by $k_B T I[X;\tilde{X}]\ln 2$ \cite{sagawa2009minimal,wolpy}, which is lower-bounded by $k_B T R(D)\ln 2$.  This is a different energetic consideration than that mentioned in Ref. \cite{tlusty2008}.


In short, $R(D)$ places a lower bound on the size of the physical substrate, on timing delays between encoding and decoding environmental input, and on the power required to maintain the sensor.


To calculate $R(D)$ given a distortion measure and probability distribution over inputs, we find the $p_{\beta}(\tilde{x}|x)$ which minimizes the rate-distortion Lagrangian, $\beta \langle d(x,\tilde{x})\rangle_{p(x,\tilde{x})} + I[X;\tilde{X}]$ using the Blahut-Arimoto algorithm \cite{Cove06a}, and calculate the resultant rate $R_{\beta}$ and expected distortion $D_{\beta}$.  As $\beta$ sweeps from $0$ (high distortion) to $\infty$ (low distortion), $R_{\beta}$ and $D_{\beta}$ parametrically trace out the rate-distortion function $R(D)$.

The rate-distortion functional also has a physical interpretation as a total energetic cost.
Distortion $D$ quantifies the food energy that the organism failed to intake from the environment; rate $r$ is correlated with the energy that the organism expended in doing so.  We can loosely think of the rate $r$ as a proxy for neuron number, so that the energy expenditure of this organism's brain is $\beta^{-1}r$, where $\beta$ is the average rate of energy use for a single neuron \cite{herculano2011scaling}.  The overall energetic cost to the organism is then $D + \beta^{-1} r$, and the fitness of an organism is some monotonically decreasing function of the organism's energetic cost.

\section{Weak universality of the rate-distortion function}
\label{sec:WU}

\begin{figure}[h]
\centering
\includegraphics[width=\textwidth]{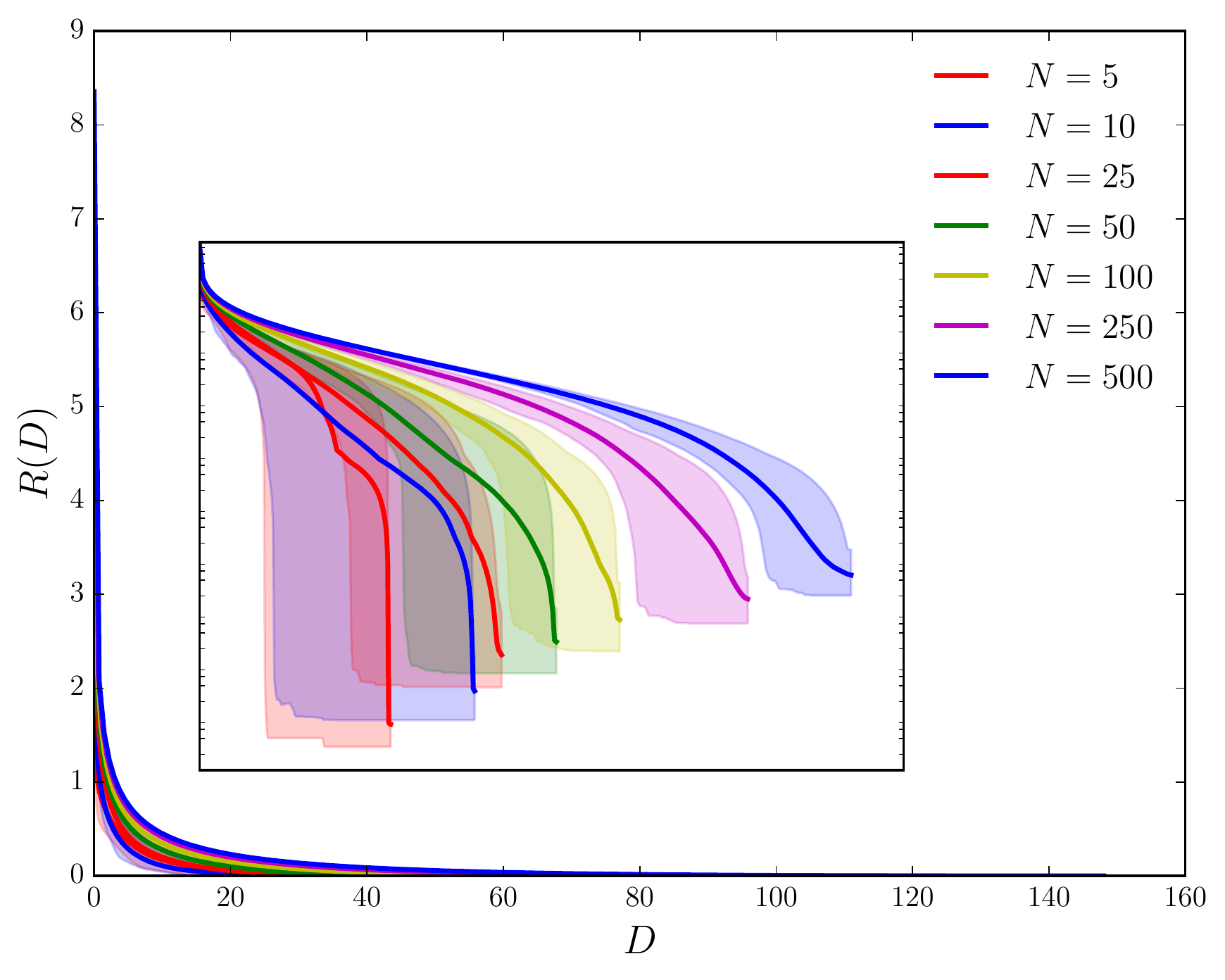}
\caption{\textbf{Convergence in probability to a single rate-distortion function.}  Each line shows the average $R(D)$ for $100$ rate-distortion functions calculated for worlds with equivalent generative models, with $\alpha=1$ and $\rho(d) = \frac{1}{\sigma d\sqrt{2\pi}} e^{-(\log d - \mu)^2/2\sigma^2}$ with $\mu \approx 3$ and $\sigma\approx 2$.  Lines denote estimates of $\bar{R}_{N}(D)$ and surrounding transparent regions show $68\%$ confidence intervals on $R(D)$ obtained by bootstrapping.  Linear interpolation is used to find $R(D)$ at desired $D$'s from the distortions at which $R(D)$ was actually calculated.  The inset is a log-log plot of the same.  The size of the $68\%$ confidence intervals appear to decrease, and the average rate-distortion function $\bar{R}_N(D)$ appears to increase, as $N$ grows larger.}
\label{fig:1}
\end{figure}

Numerical experiments shown in Fig.~\ref{fig:1} strongly suggest that, when there are many possible environmental inputs ($N\gg1$), the rate-distortion function $R(D)$ does not depend on the specific distortion measure or environmental input probabilities, but only on the distribution from which distortions were drawn, $\rho(d)$, and the distribution from which the input probabilities were drawn, characterized by concentration parameter $\alpha$.  Note that Ref. \cite{marzen2015evolution} considered the effects of a nonzero $d_\mathrm{min} = \inf \{x:\rho(x)>0\}$.  Here, we assume that $d_\mathrm{min}=0$.

We refer to the insensitivity of the rate-distortion function to the particular distortion measure and probability distribution over inputs as ``weak universality'' \footnote{``Strong universality'' would imply that the rate-distortion function was equivalent for different distortion measures and probability distribution over inputs.}.
In particular, we now argue that the rate-distortion function converges in probability to a curve which depends only on $\rho$ and $\alpha$.  Let subscripts of $R_{N,\bf{d}}(D)$ denote the number of sensory inputs $N$ and the distortion measure $\bf{d}$.  We wish to show that
\begin{equation}
\lim_{N\rightarrow\infty} \mathbb{P}(|R_{N,\bf{d}}(D)-\lim_{N\rightarrow\infty} \langle R_{N,\bf{d}}(D)\rangle_{p(\bf{d})}|\geq \epsilon) = 0~\forall~\epsilon>0. \label{eq:conj}
\end{equation}
If Eq.~\ref{eq:conj} holds, then (loosely speaking) the rate-distortion function $R_{N,\bf{d}}(D)$ depends only on $\rho$ and $\alpha$ in the large $N$ limit, \emph{even though} optimal codebooks for distortions with the same $\rho$ and $\alpha$ but different $\bf{d}$ tend to differ wildly.

To do so, we must first argue that $\lim_{N\rightarrow\infty} \langle R_{N,\bf{d}}(D)\rangle_{p(\bf{d})}$ exists.  Ref. \cite{marzen2015evolution} showed that $\langle R_{N,\bf{d}}(D)\rangle_{p(\bf{d})} \leq \log \frac{1}{\int_D^{\infty} \rho(x) dx}$ in the large $N$ limit, which implies that $\langle R_{N,\bf{d}}(D)\rangle_{p(\bf{d})}$ is bounded from above.  Simulation results suggest that $\langle R_{N,\bf{d}}(D)\rangle_{p(\bf{d})}$ is strictly increasing with $N$; see Ref.~\cite{marzen2015evolution} and Fig.~\ref{fig:1} for examples.  The monotone convergence theorem then implies that $\lim_{N\rightarrow\infty} \langle R_{N,\bf{d}}(D)\rangle_{p(\bf{d})}$ exists.  For ease, we introduce new notation: $\bar{R}_N(D) := \langle R_{N,\bf{d}}(D)\rangle_{p(\bf{d})}$ and
\begin{equation}
\bar{R}(D) := \lim_{N\rightarrow\infty} \bar{R}_N(D)
\end{equation}
where $\bar{R}_{N}(D)$ and $\bar{R}(D)$ depend on both $\rho$ and $\alpha$.

\begin{figure}[h]
\centering
\includegraphics[width=\textwidth]{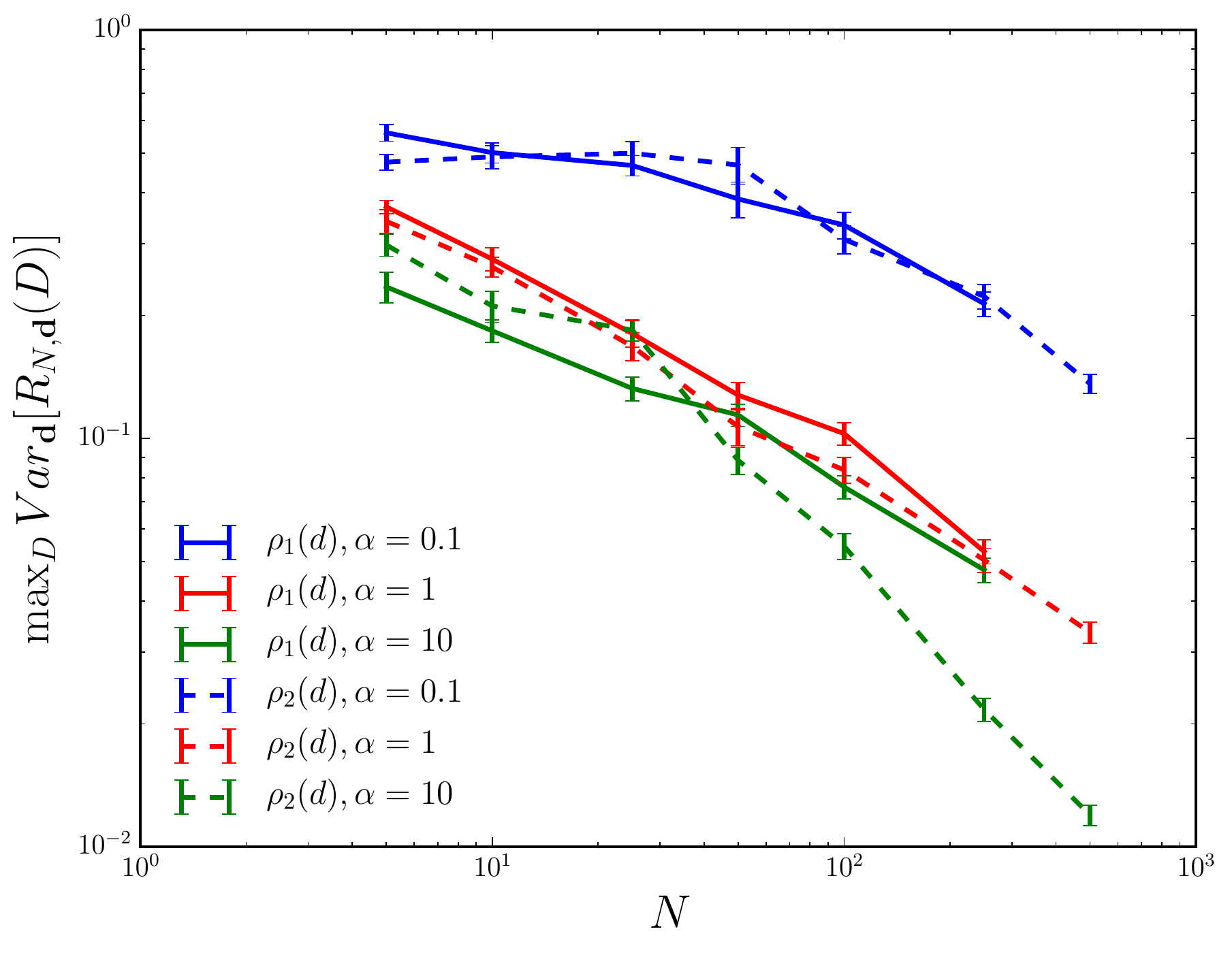}
\caption{\textbf{Fluctuations in required resources diminish as the environmental complexity increases.} Maximal variance in $R(D)$ calculated for three different world generative models at $N=5,~10,~25,~50,~100,~250,~500$.  The distortions are chosen from $\rho_1(d) = e^{-d}$ and $\rho_2(d)= \frac{1}{\sigma d\sqrt{2\pi}} e^{-(\log d - \mu)^2/2\sigma^2}$ with $\mu \approx 3$ and $\sigma\approx2$; the input probability distributions are Dirichlet with concentration parameter $\alpha$.  The y-axis values, $\max_D \text{Var}_{\bf{d}}[R_{N,\bf{d}}(D)]$, are estimated by bootstrapping given $100$ samples of the rate-distortion function at $200$ uniformly spaced distortions between $0$ and the maximum $D_{max}$. In all cases, deviations from the average case decline rapidly with $N$; this is true despite large differences in the moments of the different distributions considered.}
\label{fig:2}
\end{figure}

An application of Markov's inequality with nonnegative random variable $X = |R_{N,\bf{d}}(D)-\bar{R}(D)|$ reveals that
\begin{widetext}
\begin{eqnarray}
\mathbb{P}(|R_{N,\bf{d}}(D)-\bar{R}(D)|\geq \epsilon) &\leq& 2\bar{R}(D)\frac{\sqrt{\langle (R_{N,\bf{d}}(D)-\bar{R}_N(D))^2\rangle_{p(\bf{d})}}}{\epsilon},
\label{eq:3}
\end{eqnarray}
\end{widetext}
where we have used $\langle |R_{N,\bf{d}}(D)-\bar{R}(D)|\rangle_{p(\bf{d})}  \leq \langle R_{N,\bf{d}}(D)\rangle_{p(\bf{d})} + \bar{R}(D) \leq 2\bar{R}(D)$.
For Eq.~\ref{eq:conj} to hold, we must show that the right hand side of Eq.~\ref{eq:3} tends to $0$ as $N\rightarrow\infty$.  We argued above that $\bar{R}(D)$ was bounded.  Fig.~\ref{fig:2} suggests that $\lim_{N\rightarrow\infty} \sqrt{\langle (R_{N,\bf{d}}(D)-\bar{R}_N(D))^2\rangle_{p(\bf{d})}} = 0$ for all $D$.
Altogether, then, we have numerical evidence that Eq.~\ref{eq:conj} holds for any $D$, i.e. that $R_{N,\bf{d}}(D)$ converges in probability to $\bar{R}(D)$.

In some ways, the results presented above are unsurprising.  The rate-distortion function is a one-dimensional projection of $N(N-1)$ i.i.d. distortions, so we might expect weak universality to emerge from some particular application of the weak law of large numbers.  Indeed, we can show analytically that this happens in the two extreme limits (low and high distortion), thus providing some insight into the mechanism by which rate-distortion function converges in probability to $\bar{R}(D)$.  Proof of convergence at other distortions remains an open problem.

In the low distortion limit, there is an exact expression for $R_{\beta}$, $D_{\beta}$ in Ref. \cite{Berg71a}.
Repeated applications of the weak law of large numbers in the large $N$ limit yields
\begin{eqnarray}
D_{\beta} &=& \frac{\langle de^{-\beta d}\rangle_{\rho(d)}}{\langle e^{-\beta d}\rangle_{\rho(d)}} \\
R_{\beta} &=&\psi(N\alpha)-\psi(\alpha) -\beta D_{\beta} + \nonumber \\
&& - \log(1+N\langle e^{-\beta d}\rangle_{\rho(d)})
\end{eqnarray}
to lowest order in $N\langle e^{-\beta d}\rangle_{\rho(d)}$; the first two terms in the expansion of $R_\beta$ are equal to the expectation value of the entropy of the Dirichlet distribution with parameter $\alpha$~\cite{wolpert1995estimating}. Unlike the histogram of eigenvalues of random matrices, these expressions show that there is \emph{only} weak universality, as the moment-generating function $\langle e^{-\beta d}\rangle_{\rho(d)}$ generally uniquely specifies $\rho(d)$, and different $\alpha$ have different $\psi(N\alpha)-\psi(\alpha)$.

A similar result exists in the high-distortion limit.  With zero rate, the minimal achievable distortion  ($D_{max} = \min_{I[X;\tilde{X}]=0} \sum_{x,\tilde{x}} p(x,\tilde{x}) d(x,\tilde{x})$) is equivalent to $\min_{\tilde{x}} \sum_x p(x) d(x,\tilde{x})$.  The expected value of $\sum_x p(x) d(x,\tilde{x})$ over the ensemble of environments is $(1-\frac{1}{N}) \langle d\rangle_{\rho(d)}$ and the variance is $\frac{1}{N\alpha+1} (1-\frac{1}{N})^2 \langle d^2\rangle_{\rho(d)} - \frac{\langle d\rangle_{\rho(d)}^2}{N^2}$, as shown in the appendix.  As long as $\rho(d)$ has a finite second moment, this variance in $R(D)$ scales as $\sim \frac{\langle d^2\rangle_{\rho(d)}}{\alpha}\frac{1}{N}$, which tends to $0$ as $N$ grows larger; therefore, $D_{max}$ converges in probability to $\langle d\rangle_{\rho(d)}$ as $N$ tends to infinity.



Although the rate-distortion function appears invariant to changes in the particular distortion measure and probability distribution over inputs, near-optimal codebooks vary wildly from one environment to the next.  The statistics of near-optimal codebooks at an expected distortion $D$ are dictated by the $p(\tilde{x}|x)$ for which $\langle d(x,\tilde{x})\rangle_{p(x,\tilde{x})}\leq D$ and $I[X;\tilde{X}]$ is at a minimum \cite{Berg71a}, and numerical experiments show that the statistics of such $p(\tilde{x}|x)$ are heavily dependent on environment.

\section{Implications for biological organisms}
\label{sec:Conclusion}

Organisms can employ one of a few strategies to cope with wildly fluctuating environments.  The first strategy derives from Kelly's classical analysis of gambling, applied to phenotypic bet-hedging---that a population of organisms should develop into a range of phenotypes to maximize expected log growth rate \cite{Cove06a,donaldson2010fitness}.  Another strategy would involve delaying development of key brain regions until the organism has received strong environmental cues.

A third strategy would be to essentially ignore environmental fluctuations.  At first, this seems like a suboptimal strategy, in that a population of organisms that employ either of the two strategies listed above would have a higher log growth rate.  However, the weak universality results presented here suggest that the necessary size of sensory brain regions, the minimum possible timing delays in sensory perception, and the minimal power required to maintain sensory brain regions all depend only on coarse environmental statistics, even though optimal neural wiring fluctuates wildly from environment to environment.  In the examples discussed in the main text, these coarse environmental statistics are $\rho(d)$ and $\alpha$.
More generally, these coarse environmental statistics are the parameters specifying the distribution from which distortion measures are drawn and the distribution from which probability distributions over inputs are drawn.

In apparent agreement with these findings, environmental cues are scarce during development, and seem to have limited effect on neuron number~\cite{williams1988control}, and there are few reports of neurogenesis in mammalian sensory brain regions \cite{kaslin2008proliferation}.

Number may be fixed, but wiring is not, and there are many reports of synaptic plasticity in sensory brain regions; the particular wiring of neurons in sensory brain regions does depend on the details of environmental cues~\cite{kaas1991plasticity, buonomano1998cortical}.

If weak universality-type results mean that sensory neuron number can be largely fixed ahead of time, two questions immediately suggest themselves.
First, why do investigators find evidence of neurogenesis in non-vertebrate sensory brain regions \cite{kaslin2008proliferation}?  And second, why is there high phenotypic variability in sensory neuron number for many animals, including primates \cite{williams1988control}?

First, Ref. \cite{kaslin2008proliferation} notes that animals with substantial neurogenesis in sensory areas are also those that grow considerably postnatally, which---in our simple conception of organisms---corresponds to an increase in the possible actions $a$ taken by the organism.  Recall that one can connect the distortion measure directly to the reward function $r(x,a)$ and action policy $p(a|\tilde{x})$.  Changes in the set of actions will thus change the distortion measure  in a (possibly) more structured way than what was considered here.  That, in turn, will likely lead to an increase in the requisite sensor size, necessitating adult neurogenesis in sensory areas.  We leave a delineation of the induced structure in the distortion measure to future research.

Meanwhile, phenotypic variability is explainable within our minimal model.
Earlier, we identified the rate-distortion objective as a fitness function, implying that variability in the Lagrange multiplier $\beta$ (representative of single neuron power usage) is tightly connected to variability in the observed number of sensory neurons.  This explanation could be tested by correlating sensory neuron number with the average heat dissipation rate of single neurons in sensory regions.

Our minimal model of sensory tradeoffs in biological organisms lead to new questions at the intersection of random matrix theory, information theory, and sensory processing.  Extensions of this approach---to distortion measures that change as the animal grows, or to distortion measures and probability distributions over inputs with more structure---may predict and mathematically explain other observed similarities and differences between species.



\section*{Acknowledgments}

The authors thank C. Hillar, J. P. Crutchfield, and anonymous referees,
for helpful discussions and the Santa Fe Institute for its hospitality during
visits. SM was funded by a
National Science Foundation Graduate Student Research Fellowship and the U.C.
Berkeley Chancellor's Fellowship.

\bibliography{rmrd2}

\appendix 

\begin{widetext}

\section{Weak universality in the low- and high-distortion limits}
\label{app:B}


First we tackle the low-distortion limit.  When $p_{\beta}(\tilde{x})$ has full support, an exact expression exists for $R_{\beta}$ and $D_{\beta}$ from \cite{Berg71a}.   Let $\vec{p}(x)$ be a vector of input probabilities $p(x)$, let $\bf{d}$ be the distortion matrix, and let $Q_{x,\tilde{x}}=e^{-\beta d(x,\tilde{x})}$.  Then
\begin{eqnarray}
R_{\beta} &=& -\beta D_{\beta} +H[X] + \vec{p}(x)^{\top} \log (Q^{-1}\vec{1}) \\
D_{\beta} &=& [\frac{\vec{p}(x)}{Q^{-1}\vec{1}}]^{\top} Q^{-1} (\textbf{d} \odot Q) (Q^{-1}\vec{1}).
\end{eqnarray}
When $\beta$ is sufficiently large, then the entries of $Q-I$ are much smaller than $1$ with high probability, suggesting the expansion
\begin{eqnarray*}
Q &=& I + (Q-I) \\
Q^{-1} &=& \left(I + (Q-I)\right)^{-1} \\
&=& \sum_{m=0}^{\infty} (-1)^m (Q-I)^m. \\
\end{eqnarray*}
By the weak law of large numbers, $(Q-I)\vec{1}$ is highly concentrated around $(N-1)\langle e^{-\beta d}\rangle_{\rho(d)} \vec{1}$ as long as the probability density function for $e^{-\beta d}$ has finite variance, so that
\begin{equation*}
(Q-I)^m \vec{1} \approx \left((N-1)\langle e^{-\beta d}\rangle_{\rho(d)}\right)^m \vec{1}
\end{equation*}
showing that
\begin{eqnarray*}
Q^{-1}\vec{1} &=& \sum_{m=0}^{\infty} \left(-(N-1)\langle e^{-\beta d}\rangle_{\rho(d)}\right)^m \vec{1} \\
&\approx& (1+(N-1)\langle e^{-\beta d}\rangle_{\rho(d)})^{-1}\vec{1}.
\end{eqnarray*}
Then we find that
\begin{eqnarray*}
\vec{p}(x)^{\top} \log (Q^{-1}\vec{1}) &\approx & \sum_x p(x) \log (1+(N-1)\langle e^{-\beta d}\rangle_{\rho(d)})^{-1} \\
&=& -\log \left(1+(N-1)\langle e^{-\beta d}\rangle_{\rho(d)}\right)
\end{eqnarray*}
so that
\begin{equation}
R_{\beta} = -\beta D_{\beta} + H[X] - \log \left(1+(N-1)\langle e^{-\beta d}\rangle_{\rho(d)}\right).
\label{eq:RbetaU}
\end{equation}
Similar manipulations, again based on the weak law of large numbers, reveal that
\begin{eqnarray}
D_{\beta} &\approx & \frac{(N-1)\langle de^{-\beta d}\rangle_{\rho(d)}}{1+(N-1)\langle e^{-\beta d} \rangle_{\rho(d)}}.
\label{eq:DbetaU}
\end{eqnarray}
When the probability distribution over inputs is drawn from a Dirichlet distribution with concentration parameter $\alpha$, then $H[X]$ is very peaked around $\psi(N\alpha)-\psi(\alpha)$ with corrections of $O(1/N)$ \cite{nemenman2001entropy}, yielding
\begin{equation}
R_{\beta} \approx -\beta D_{\beta} + \psi(N\alpha)-\psi(\alpha) - \log \left(1+(N-1)\langle e^{-\beta d}\rangle_{\rho(d)}\right).
\label{eq:RbetaU2}
\end{equation}
Thus $D_{\beta}$ and $R_{\beta}$ are independent of the particular distortion matrix and dependent only on $\rho(d),~\alpha$.
Unlike the histogram of eigenvalues of random matrices, there is \emph{only} weak universality, as the moment-generating function $\langle e^{-\beta d}\rangle_{\rho(d)}$ specifies $\rho(d)$ and different $\alpha$ have different $\psi(N\alpha)-\psi(\alpha)$.  Note that this formula holds only when $p_{\beta}(\tilde{x})$ has full support, or (roughly speaking) when $N\langle e^{-\beta d}\rangle_{\rho(d)}\ll 1$.

Next, we tackle the high-distortion limit, i.e. find weak universality in $D_{max}$.  We look to show that $\min_{\tilde{x}}\sum_x p(x) d(x,\tilde{x})$ converges in probability $\langle d\rangle_{\rho(d)}$, and so we look to show that the expected value of $\min_{\tilde{x}}\sum_x p(x) d(x,\tilde{x})$ over worlds with the same $\rho(d),~\alpha$ converges in probability to $\langle d\rangle_{\rho(d)}$.  For any $\tilde{x}$ we have
\begin{eqnarray*}
\langle \sum_x p(x) d(x,\tilde{x})\rangle_{p(\vec{p}(x),\bf{d})} &=& (N-1) \left(\frac{\alpha}{N\alpha}\right) \langle d\rangle_{\rho(d)} \\
&=& (1-\frac{1}{N}) \langle d\rangle_{\rho(d)}
\end{eqnarray*}
while the variance of $\sum_x p(x) d(x,\tilde{x})$ is
\begin{eqnarray*}
\langle \left(\sum_x p(x) d(x,\tilde{x})-\langle \sum_x p(x) d(x,\tilde{x})\rangle_{p(\vec{p}(x),\bf{d})}\right)^2\rangle_{p(\vec{p}(x),\bf{d})} &=& \langle (\sum_x p(x) d(x,\tilde{x}))^2 \rangle_{p(\vec{p}(x),\bf{d})} - \langle \sum_x p(x) d(x,\tilde{x})\rangle_{p(\vec{p}(x),\bf{d})}^2 \\
&=& \langle \sum_{x,x'} p(x) p(x') d(x,\tilde{x}) d(x',\tilde{x}) \rangle_{p(\vec{p}(x),\bf{d})} - (1-\frac{1}{N})^2 \langle d\rangle_{\rho(d)}^2 \\
&=& \langle \sum_{x} p(x)^2 d(x,\tilde{x})^2 \rangle_{p(\vec{p}(x),\bf{d})} + \sum_{x\neq x'} \langle p(x) p(x') d(x,\tilde{x}) d(x',\tilde{x})\rangle_{p(\bf{d})} \nonumber \\
&& - (1-\frac{1}{N})^2 \langle d\rangle_{\rho(d)}^2\\
&=& (N-1) \langle p(x)^2 \rangle_{p(\vec{p}(x))} \langle d^2\rangle_{\rho(d)} + (N-1) (N-2) \langle p(x) p(x')\rangle \langle d\rangle_{\rho(d)}^2\nonumber \\
&& - (1-\frac{1}{N})^2 \langle d\rangle_{\rho(d)}^2
\end{eqnarray*}
where $x\neq x'$ in the second term.
The first of these terms is relatively easy to evaluate using the fact that if $x_1,\ldots,x_N$ are drawn from a Dirichlet distribution with concentration parameter $\alpha$, then $\rho(x_1) = B(x_1;\alpha,(N-1)\alpha)$ and $\rho(x_2|x_1) = B(\frac{x_2}{1-x_1};\alpha,(N-2)\alpha)$ \cite{nemenman2001entropy}.  As such, we find that
\begin{eqnarray*}
\langle p(x)^2\rangle_{p(\vec{p}(x))} &=& \int_0^1 p(x_1)^2 B(p(x_1);\alpha,N\alpha) dp(x_1) \\
&=& (1-\frac{1}{N})^2\frac{1}{N\alpha+1}.
\end{eqnarray*}
The second term is evaluated by noticing
\begin{eqnarray*}
\langle p(x) p(x')\rangle_{p(\vec{p}(x))} &=& \int \Prob(p(x_1),\ldots,p(x_N)) p(x_1) p(x_2) dp(x_1)\ldots dp(x_N) \\
&=& \int \int \Prob(p(x_1),p(x_2)) p(x_1) p(x_2) dp(x_1) dp(x_2) \\
&=& \int_0^1 \int_0^1 p(x_1) p(x_2) B(p(x_1);\alpha,(N-1)\alpha) B(\frac{p(x_2)}{1-p(x_1)};\alpha,(N-2)\alpha) dp(x_1) dp(x_2).
\end{eqnarray*}
After some algebra, we find that $\langle p(x) p(x')\rangle = \frac{1}{N(N-1)}$ for $x\neq x'$, and so
\begin{eqnarray*}
\langle \left(\sum_x p(x) d(x,\tilde{x})-\langle \sum_x p(x) d(x,\tilde{x})\rangle_{p(\vec{p}(x),\bf{d})}\right)^2\rangle_{p(\vec{p}(x),\bf{d})} &=& (1-\frac{2}{N}) \langle d\rangle_{\rho(d)}^2 + (1-\frac{1}{N})^2 \frac{\langle d^2\rangle_{\rho(d)}}{N\alpha+1} - (1-\frac{2}{N}+\frac{1}{N^2}) \langle d\rangle_{\rho(d)}^2 \\
&=& \frac{1}{N\alpha+1} (1-\frac{1}{N})^2 \langle d^2\rangle_{\rho(d)} - \frac{\langle d\rangle_{\rho(d)}^2}{N^2}.
\end{eqnarray*}
In the large $N$ limit, we have $\langle \left(\sum_x p(x) d(x,\tilde{x})-\langle \sum_x p(x) d(x,\tilde{x})\rangle_{p(\vec{p}(x),\bf{d})}\right)^2\rangle_{p(\vec{p}(x),\bf{d})} \sim \frac{\langle d^2\rangle_{\rho(d)}}{N\alpha}$.  Chebyshev's inequality implies that $\sum_x p(x) d(x,\tilde{x})$ tends to $\langle d\rangle_{\rho(d)}$ in probability as $N\rightarrow \infty$ for all $\tilde{x}$, and so $D_{max} = \min_{\tilde{x}}\sum_x p(x) d(x,\tilde{x})$ converges in probability to $\langle d\rangle_{\rho(d)}$ in probability as $N\rightarrow\infty$. 

\end{widetext}

\end{document}